# Spin-orbit-enhanced correlation sensitivity and anomalous magnetic response in non-dimerized $4d^4$ ilmenite $CdRuO_3$

Yuya Haraguchi†, Hayato Yatsuzuka, and Hiroko Aruga Katori
Department of Applied Physics and Chemical Engineering, Tokyo University of Agriculture and Technology, Koganei, Tokyo 184-8588, Japan
†Corresponding author: chiyuya3@go.tuat.ac.jp

We report the synthesis and physical properties of $CdRuO_3$, a nominal $Ru^{4+}$ $4d^4$ ilmenite with edge-sharing $RuO_6$ honeycomb layers. Powder x-ray diffraction establishes a crystallographically non-dimerized $R\overline{3}$ structure with equivalent Ru–Ru bonds and a strongly distorted $RuO_6$ environment. The compacted-pellet resistivity is nonmetallic but non-Arrhenius, while the heat capacity contains a finite residual linear term. Matched nonmagnetic calculations show that PBE+U without spin–orbit coupling remains metallic or semimetallic up to $U_{\mathrm{eff}}$ = 3 eV, whereas PBE+SOC+$U$ exhibits a strong $U_{\mathrm{eff}}$ dependence and opens a direct gap of approximately 55 meV at Γ for $U_{\mathrm{eff}}$ = 2.5 eV. Spin–orbit coupling therefore markedly enhances the correlation sensitivity of the non-cubic Ru $t_{2g}$ manifold. After subtraction of a dilute Curie–Weiss defect contribution, the susceptibility remains weakly nonmonotonic and is inconsistent with both an ordinary Pauli response and independent spin-only $S = 1$ moments. $CdRuO_3$ thus realizes the non-dimerized structural branch predicted for ruthenium ilmenites, but not a simple robust multiorbital metal.

## A. Introduction

The electronic state of a nominal $4d^4$ ion is governed by competing local Coulomb interaction, Hund coupling, spin-orbit coupling, non-cubic crystal fields, and intersite hopping [1]. In the localized octahedral limit, low-spin $Ru^{4+}$ has an $S = 1$ and $L_{\mathrm{eff}} = 1$ manifold, from which spin-orbit coupling can generate a nonmagnetic singlet and low-lying magnetic excitations [2]. Exchange-driven magnetic states emerging from such a $d^4$ manifold have also been analyzed theoretically [3]. Increasing intersite hopping instead favors an itinerant Ru $t_{2g}$ state, while strong non-cubic crystal fields split and mix both the orbital and ideal $J_{\mathrm{eff}}$ sectors. Determining which limit applies in an actual $4d^4$ oxide therefore requires its lattice structure, charge response, magnetic response, and electronic structure to be considered together.

Honeycomb ruthenates illustrate this competition particularly clearly. $Li_2RuO_3$ undergoes a high-temperature structural transition [4], which was subsequently associated with molecular-orbital formation on short Ru-Ru bonds [5]. A dimerized superstructure driven by orbital degeneracy and covalent bonding was proposed theoretically [6], while structural and electronic studies have also identified valence-bond-like behavior [7]. $Ag_3LiRu_2O_6$ provides a non-dimerized comparison, where magnetization, NMR, μSR, inelastic neutron scattering, and quantum-chemistry calculations support competing spin-orbital singlet states [8]. $Na_2RuO_3$ remains more ambiguous: antiferromagnetic insulating behavior has been reported [9], whereas thermodynamic, magnetic, and photoemission measurements have also supported a correlated-metal interpretation [10]. Suppression of crystallographic Ru-Ru dimerization therefore does not uniquely determine whether the remaining $4d^4$ state is localized or itinerant.

A concrete structural and electronic prediction for ruthenium ilmenites was provided by the first-principles comparison of $MgRuO_3$ and $CdRuO_3$ [13]. In $MgRuO_3$, structural optimization produced a pronounced shortening of one of the three Ru-Ru bonds in the honeycomb layer. The $t_{2g}$ orbital associated with that bond was split into bonding and antibonding bands, while the remaining two $t_{2g}$ orbitals formed the Fermi surfaces. $CdRuO_3$ was predicted to lie on the opposite structural branch. Its optimized structure remained close to $R\overline{3}$, the energy was minimized for equivalent Ru-Ru bonds, and no secondary energy minimum appeared upon contraction or elongation of a selected Ru-Ru bond. All three Ru $t_{2g}$ orbitals consequently contributed almost equally to metallic Fermi surfaces [13]. The Mg-Cd comparison therefore proposed an A-site-size-controlled relationship among Ru-Ru dimerization, orbital selectivity, and itinerancy.

This prediction gives $CdRuO_3$ a specific experimental role. If the predicted non-dimerized multiorbital state were robust, one would expect equivalent nearest-neighbor Ru-Ru bonds, metallic or semimetallic transport, and a predominantly Pauli-like magnetic susceptibility. A conventional localized-spin limit would instead produce a pronounced $S = 1$ Curie-Weiss response. A third possibility arises because equivalent Ru-Ru bonds do not imply a locally cubic Ru environment. The ilmenite coordination can preserve the average $R\overline{3}$ honeycomb symmetry while retaining a strong trigonal or more general non-cubic $RuO_6$ distortion. Such a distortion can split and mix the Ru $t_{2g}$ states and strongly modify their response to both spin-orbit coupling and local Coulomb interaction. Experiments on $MgIrO_3$ and $ZnIrO_3$ established strong magnetic anisotropy in related honeycomb ilmenites [11], while $CdIrO_3$ demonstrated that a large local trigonal distortion can substantially alter magnetic interactions within the same structural framework [12].

The present study addresses three specific questions. First, does synthesized $CdRuO_3$ realize the crystallographically non-dimerized structural branch predicted for the larger-A-site ruthenium ilmenite? Second, is the predicted multiorbital metallic state robust against moderate local Coulomb interaction, and how does spin-orbit coupling alter that response? Third, is the measured susceptibility consistent with an ordinary Pauli metal or independent spin-only $S = 1$ $Ru^{4+}$ moments? We combine topochemical synthesis, powder x-ray diffraction, transport, heat capacity, magnetization, matched spin-unpolarized PBE+$U$ calculations without spin-orbit coupling, and nonmagnetic PBE+SOC+$U$ calculations. The experimental structure confirms the non-dimerized branch, but the calculations show that spin-orbit coupling strongly enhances the sensitivity of the Ru $t_{2g}$ bands to $U_{eff}$, while the defect-corrected susceptibility departs from both simple magnetic limits.

## B. Experimental Procedure

Polycrystalline $CdRuO_3$ was synthesized by topochemical cation exchange starting from a layered rocksalt-type $Li_2RuO_3$ precursor. The $Li_2RuO_3$ precursor was prepared by a conventional solid-state reaction from $Li_2CO_3$ and $RuO_2$, with repeated grinding and annealing. The phase purity of the precursor was confirmed by powder x-ray diffraction.

For the ion-exchange step, $Li_2RuO_3$ was mixed with $CdSO_4$ and a $CdCl_2$-NaCl eutectic salt, pelletized, sealed in an Ar-filled tube, and annealed at 350 °C for 100 h. This low-temperature reaction was designed to enable cation exchange while preserving the transition-metal-oxide framework. After the reaction, the product was thoroughly washed with distilled water to remove soluble byproducts and residual salts and was then dried. Powder x-ray diffraction showed no detectable precursor peaks after ion exchange within the sensitivity of the present laboratory powder x-ray diffraction measurement.

The low-temperature salt-mediated route follows the synthetic strategy previously used to access metastable ilmenite-type honeycomb oxides. $MgIrO_3$ and $ZnIrO_3$ were synthesized through a low-temperature topochemical reactions [11]. $CdIrO_3$ was subsequently stabilized using a related low-temperature topochemical approach [12]. The present synthesis extends this strategy to a nominal $Ru^{4+}$ $4d^4$ ilmenite by replacing Li with Cd under conditions designed to preserve the connectivity of the Ru-O framework.

The cation stoichiometry of $CdRuO_3$ was evaluated by energy-dispersive x-ray spectroscopy using JSM-IT 100. The measured ratio was Cd:Ru = 1.01(2):0.99(3), consistent with a 1:1 cation composition within uncertainty. Together with the absence of detectable $Li_2RuO_3$ precursor reflections in the powder x-ray diffraction pattern, this result argues against a substantial residual precursor phase or large Li-for-Cd substitution. Since Li and oxygen cannot be quantified by EDS, the $Ru^{4+}$ $4d^4$ configuration discussed below is nominal and is based on $Cd^{2+}Ru^{4+}O_3$ charge balance and the approximately ideal Cd/Ru cation stoichiometry. Trace Li-related defects or oxygen nonstoichiometry cannot be excluded by the present composition analysis.

Powder x-ray diffraction data were collected at room temperature using Cu $K\alpha$ radiation in Bragg-Brentano geometry. Rietveld refinements were performed using an ilmenite structural model with space group $R\overline{3}$, No. 148, in the hexagonal setting. The refinements yielded lattice parameters, atomic coordinates, and isotropic displacement parameters.

$R_{wp}$ is the weighted-profile $R$ factor and quantifies the agreement between the observed and calculated diffraction profiles. The goodness-of-fit parameter is defined as $S = R_{wp}/R_e$, where $R_e$ is the statistically expected $R$ factor.

Representative Ru-O distances, O-Ru-O angles, Ru-O-Ru angles, and Ru-Ru distances derived from the refined structure are summarized in Supplementary Table S1. Because oxygen positions are refined from laboratory powder x-ray diffraction data in a compound containing heavy Cd and Ru atoms, quantitative measures of the $RuO_6$ distortion are treated as structural estimates. This limitation applies to oxygen-derived distortion parameters, not to the crystallographic absence of a $Li_2RuO_3$-type short Ru-Ru dimer bond. The Ru bond-valence sum was calculated from the refined Ru-O distances using $V(\mathrm{Ru}) = \Sigma_i \exp[(R_0 - d_i)/B]$. The bond-valence formalism and tabulated parameters were established through systematic analysis of inorganic crystal structures [14] and were later extended in a broader set of bond-valence parameters for solids [15]. For $Ru^{4+}$-O, we used $R_0 = 1.834$ Å and $B = 0.37$ Å.

Magnetic susceptibility was measured using a Magnetic Property Measurement System (MPMS; Quantum Design) under a dc field of 1 T down to 2 K. The susceptibility reported here was measured under the field-cooled protocol. A dedicated paired ZFC–FC comparison, particularly at lower applied fields, was not performed. Accordingly, weak low-field magnetic irreversibility was not assessed in the present study. The low-temperature Curie-like upturn below 20 K was treated as a dilute impurity contribution and fitted with a Curie-Weiss-type term. This fitted contribution was subtracted from the measured susceptibility to obtain the defect-corrected susceptibility used in the following analysis. The susceptibility was not corrected for core diamagnetism, whose ordinary ionic contribution is small compared with the $10^{-3}$ $cm^3$ $mol^{-1}$ susceptibility scale discussed below. Magnetization curves were measured at 2 K up to 7 T.

**Table 1**. Crystallographic parameters for $CdRuO_3$ (space group: $R\overline{3}$) determined from powder x-ray diffraction experiments. The obtained trigonal lattice parameters are $a = 5.3276(3)$ Å and $c = 14.8585(5)$ Å. $B$ is the atomic displacement parameter.

| atom | site | $x$ | $y$ | $z$ | $B$(Å$^2$) |
|---|---|---|---|---|---|
| Cd | 6$c$ | 0 | 0 | 0.36886(6) | 0.15 |
| Ru | 6$c$ | 0 | 0 | 0.16610(6) | 0.49 |
| O | 18$f$ | 0.3660(8) | 0.0065(94) | 0.1062(2) | 1.30 |

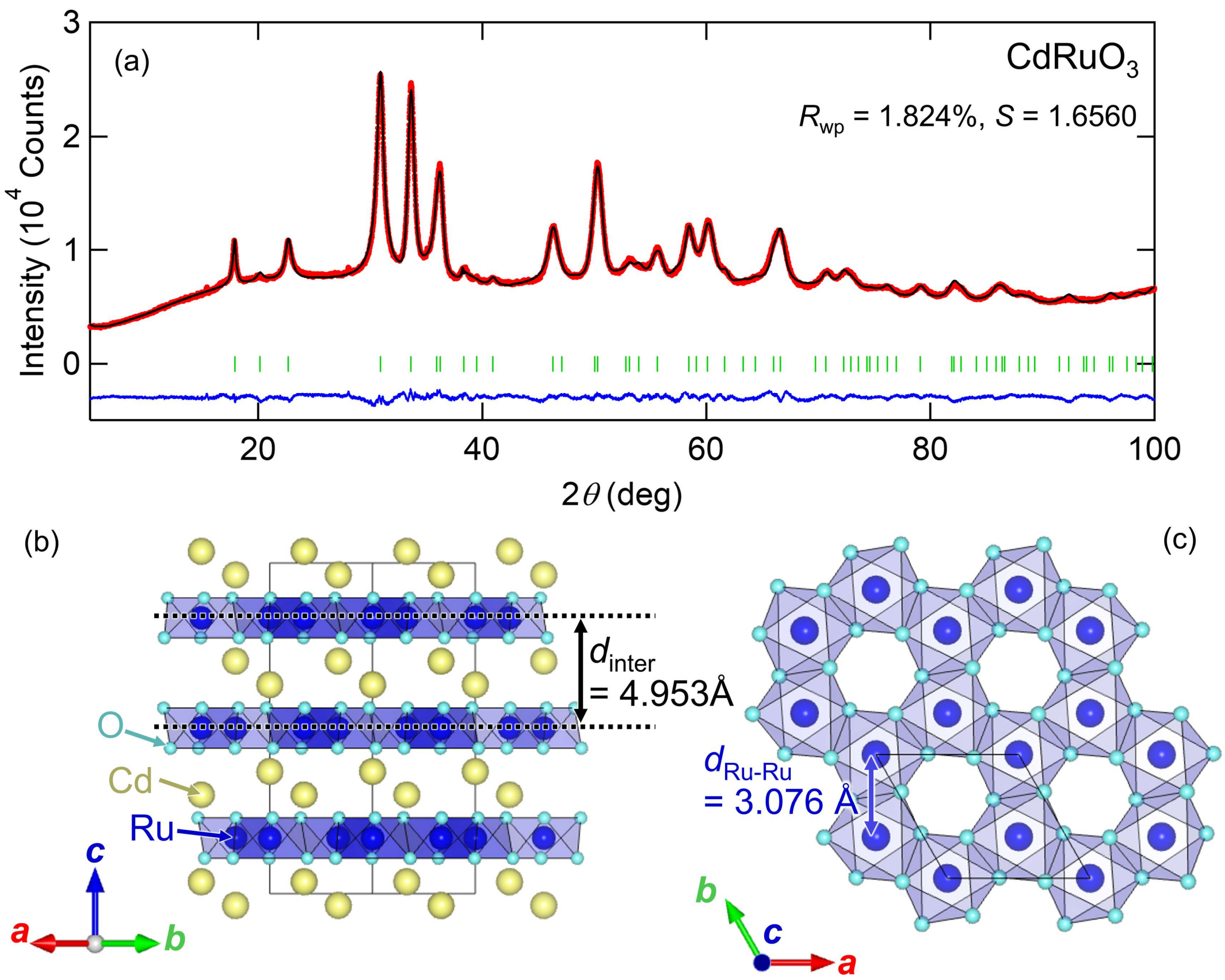

**Figure 1.** (a) Powder x-ray diffraction pattern and Rietveld refinement using the ilmenite model with space group $\boldsymbol{R\bar{3}}$. Red symbols, black line, and blue line denote observed intensity, calculated profile, and difference, respectively. Green ticks indicate calculated Bragg positions. (b) Crystal structure of $CdRuO_3$, showing alternating $CdO_6$ and $RuO_6$ layers. (c) Top view of the $RuO_6$ honeycomb layer. The nearest-neighbor Ru-Ru distance is 3.076 Å, and no crystallographically resolved short Ru-Ru dimer bond is observed. The crystal structure was visualized using VESTA [23].

Specific heat was measured using a Physical Property Measurement System (PPMS; Quantum Design) by the relaxation method. The low-temperature data were fitted using $C(T) = \gamma T + \beta T^3$. The data were collected at independently stabilized temperature points in a cooling sequence. The fitted $\gamma$ is treated as a residual linear heat-capacity coefficient and is not assumed a priori to be purely electronic.

For transport measurements, polycrystalline pellets were compacted under 2 GPa to improve grain connectivity. The pellets were then released to ambient pressure, and electrical resistance was measured using our custom-made apparatus at ambient pressure by a standard four-probe method from low temperature to room temperature. Since the sample is a compacted polycrystalline pellet, we report the apparent resistivity and focus on its temperature dependence rather than extracting a unique bulk activation gap.

First-principles calculations were performed using the Quantum ESPRESSO package. Its original modular implementation was described in Ref. [16], and its later computational capabilities were summarized in Ref. [17]. The exchange-correlation functional was the Perdew-Burke-Ernzerhof generalized-gradient approximation [18]. Scalar-relativistic and fully relativistic projector-augmented-wave pseudopotentials from the same PSLibrary family were used for calculations without and with spin-orbit coupling, respectively [19].

The calculations used the experimental $R\bar{3}$ structure represented by a primitive rhombohedral cell containing 10

atoms. The kinetic-energy cutoff for the wave functions was set to 52 Ry, and that for the charge density was set to 468 Ry. Brillouin-zone integrations employed Monkhorst-Pack meshes [20].

The self-consistent calculations used an $18 \times 18 \times 18$ $k$-point mesh, Gaussian smearing with a width of 0.001 Ry for metallic or semimetallic states, and a self-consistency threshold of $1 \times 10^{-9}$ Ry. Selected calculations were repeated using a $20 \times 20 \times 20$ mesh to test $k$-point convergence. The number of Kohn-Sham states was 92 without spin-orbit coupling and 184 in the spinor calculations.

Two matched nonmagnetic calculation series were performed using the same experimental structure:

(i) spin-unpolarized PBE+$U$ without spin-orbit coupling;
(ii) nonmagnetic noncollinear PBE+SOC+$U$.

$U_{\mathrm{eff}}$ denotes the effective on-site Hubbard interaction in the rotationally invariant Dudarev scheme [21]. Although effective Hubbard parameters can in principle be estimated using linear-response calculations [22], $U_{\mathrm{eff}}$ was treated here as a control parameter. The no-SOC series was calculated for $U_{\mathrm{eff}}$ = 0, 1, 2, 2.5, and 3 eV. The SOC series was calculated for $U_{\mathrm{eff}}$ = 0, 1, 1.5, 1.75, 2.0, 2.1, 2.3, 2.5, 2.7, and 3.0 eV.

The signed indirect gap was evaluated as $\Delta = \min_k E_{\mathrm{N+1}}(k) - \max_k E_{\mathrm{N}}(k)$. Here, $E_{\mathrm{VBM}}$ is the maximum energy of the nominal highest occupied band over all $k$ points on a uniform $k$-point mesh, and $E_{\mathrm{CBM}}$ is the minimum energy of the nominal lowest unoccupied band over the same mesh. The corresponding band indices were determined from the total electron count and the calculated occupations. A positive $\Delta$ indicates an insulating gap, whereas a negative $\Delta$ indicates an indirect overlap between the nominal valence and conduction bands. Values close to zero represent the gap-closing condition within the numerical resolution. When $E_{\mathrm{VBM}}$ and $E_{\mathrm{CBM}}$ occur at the same k point, the gap is classified as direct; otherwise, it is indirect. High-symmetry band paths were used only for visualization and not for determining $\Delta$.

The no-SOC calculations examine the spin-unpolarized solution within the experimental structural symmetry. Possible lower-symmetry orbital-polarized solutions were not systematically explored, and the calculations are not used to determine a unique orbital-ordering pattern or magnetic ground state.

Crystal structures were visualized using VESTA [23].

## C. Results

### 1. Crystal structure

Figure 1(a) shows the powder x-ray diffraction pattern and Rietveld refinement of $CdRuO_3$. The diffraction pattern is well described by an ilmenite-type structure with space group $R\overline{3}$. The refinement gives $a$ = 5.3276(3) Å and $c$ = 14.8585(5) Å, with $R_{\mathrm{wp}}$ = 1.824% and $S$ = 1.6560. No reflections attributable to the $Li_2RuO_3$ precursor are detected within the experimental sensitivity. The refined structural parameters are listed in Table 1. Cd and Ru occupy 6$c$ sites, while O occupies an 18$f$ site. The structure consists of alternating $CdO_6$ and $RuO_6$ octahedral layers stacked along the $c$ axis. Within each $RuO_6$ layer, edge-sharing octahedra form a honeycomb network, as shown in Fig. 1(b,c).

The interlayer distance between $RuO_6$ honeycomb sheets is approximately $c/3$ = 4.953 Å. The in-plane nearest-neighbor Ru-Ru distance is 3.076 Å. This distance is substantially longer than the approximately 2.57 Å short Ru-Ru dimer bond reported for the low-temperature dimerized phase of $Li_2RuO_3$ [5]. All nearest-neighbor Ru-Ru links are crystallographically equivalent in the refined $R\overline{3}$ structure. Here, “non-dimerized” refers specifically to the absence of crystallographically resolved long-range Ru-Ru bond disproportionation in the average structure. It does not exclude possible local or dynamic bond disproportionation, which would require pair-distribution-function analysis, EXAFS, electron diffraction, Raman spectroscopy, or another local-structure-sensitive probe.

At the same time, the refined $RuO_6$ environment is far from ideal. The refined oxygen positions give a bond-angle variance of 150.10 deg$^2$, corresponding to an RMS angular deviation of approximately 12.25° from an ideal octahedron. This large distortion indicates a substantial non-cubic crystal-field component acting on the Ru $t_{2g}$ manifold. Because this distortion is quantified using oxygen coordinates refined from laboratory powder x-ray diffraction, the numerical distortion parameter should be regarded as an estimate. The robust structural conclusion is that $CdRuO_3$ forms a crystallographically non-dimerized Ru honeycomb layer with a strongly distorted $RuO_6$ environment.

The bond-angle variance was calculated from the 12 cis O-Ru-O angles using $\sigma^2 = \Sigma_i(\theta_i - 90°)^2/(m - 1)$, with $m$ = 12. Using the refined Ru-O distance, the Ru bond-valence sums are 3.828. This value supports the nominal $Ru^{4+}$ $4d^4$ description. Because the bond-valence sum depends on the refined oxygen coordinates and the chosen bond-valence parameters, it is used here as a structural consistency check rather than as an independent determination of oxygen stoichiometry or Ru valence. The corresponding bond-length variance and selected bond angles are given in Supplementary Table S1.

We also compared the experimentally refined structure with the structure obtained by first-principles structural relaxation. The relaxed structure differs from the Rietveld structure, particularly in the oxygen coordinates and the resulting $RuO_6$ distortion. This comparison is used only as a reference for the ideal periodic PBE+SOC structure, because the experimental oxygen positions are obtained from room-temperature powder x-ray diffraction, and the relaxed structure represents a zero-temperature calculation within a specific functional. In the following analysis, the experimental structure is used as the primary structural basis for discussing the measured physical properties.

The cation stoichiometry was evaluated by energy-dispersive x-ray spectroscopy. The measured Cd:Ru ratio is 1.01(2):0.99(3), essentially identical to the ideal 1:1 ratio within uncertainty. This result places a strong constraint on possible residual Li-containing phases or incomplete cation exchange. A substantial amount of unreacted $Li_2RuO_3$ precursor would be expected to produce detectable precursor

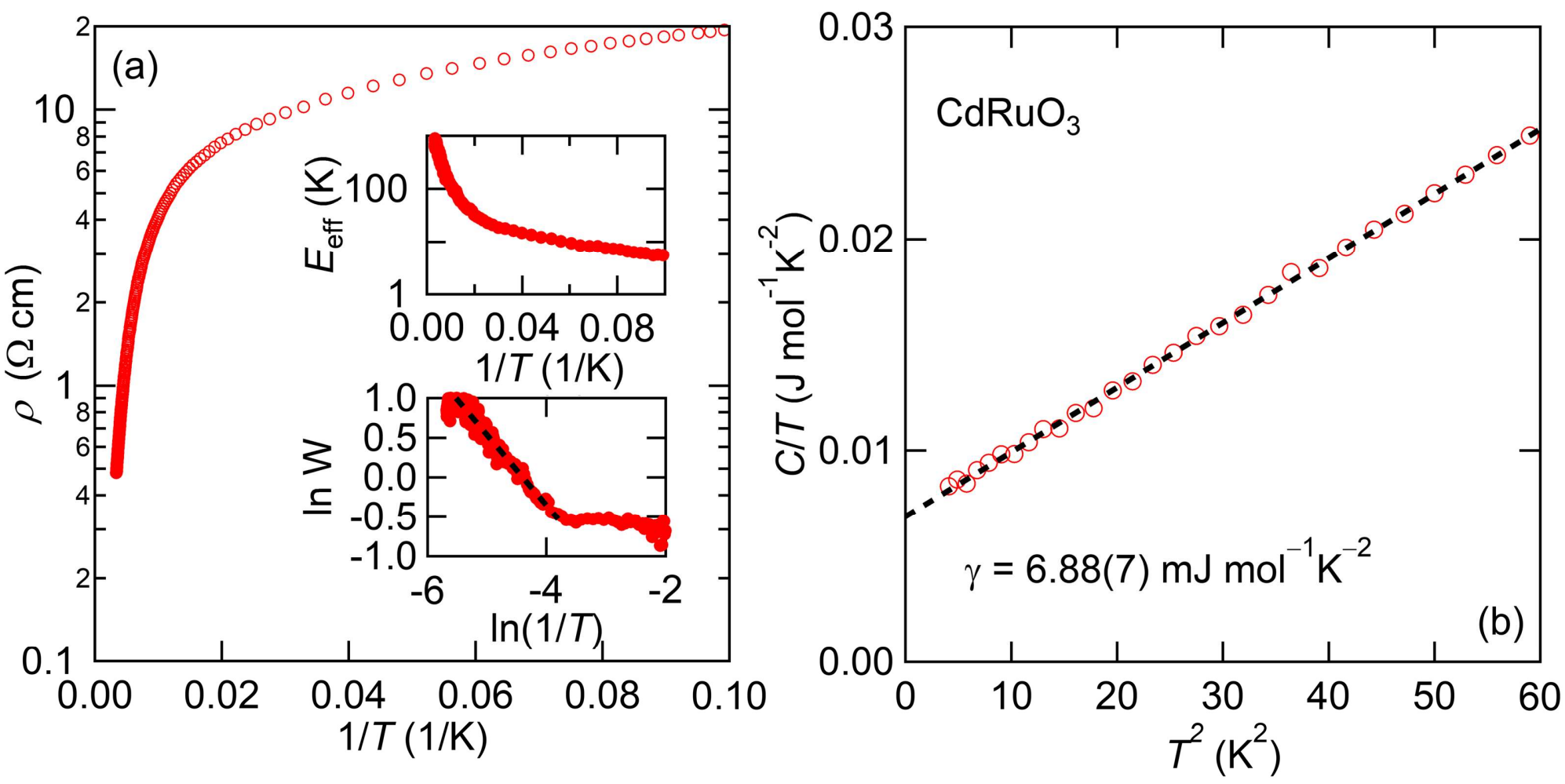

**Fig. 2** (a) Temperature dependence of apparent resistivity measured on a compacted polycrystalline pellet. Insets show the local activation scale $E_{\mathrm{eff}}(T)/k_{\mathrm{B}}$ = d(ln ρ)/d(1/$T$) and the Zabrodskii function $W(T)$ = -d(ln ρ)/d(ln $T$). $E_{\mathrm{eff}}(T)$ is strongly temperature dependent, and the Zabrodskii analysis does not reveal a robust constant hopping exponent over an extended temperature range. The pellet transport therefore cannot be assigned to a single Arrhenius gap or a conventional variable-range-hopping mechanism. Because the measurement is on a compacted pellet, the result is described as apparent resistivity rather than as a uniquely determined bulk insulating gap. (b) Low-temperature heat capacity plotted as $C/T$ versus $T^2$. The linear fit gives γ = 6.88(7) mJ mol$^{-1}$ K$^{-2}$. This finite residual linear coefficient indicates a residual low-temperature heat-capacity contribution, but it is not assigned uniquely to itinerant electronic quasiparticles.

peaks in powder x-ray diffraction or a deviation from the ideal Cd:Ru ratio. Similarly, large Li-for-Cd substitution in the Cd layer would reduce the measured Cd:Ru ratio. Neither behavior is observed. The physical properties discussed below are therefore assigned primarily to $CdRuO_3$ rather than to a significant residual precursor phase. However, trace Li-related defects or oxygen nonstoichiometry are not quantified by EDS and may contribute to the low-temperature Curie-like tail.

## 2. Resistivity and heat capacity

Figure 2(a) shows the temperature dependence of the apparent resistivity $\rho$ measured on a compacted polycrystalline pellet. The apparent resistivity increases with decreasing temperature, indicating nonmetallic pellet transport. The data, however, do not follow a single Arrhenius law over an extended temperature range. A local activation-scale analysis, $E_{\mathrm{eff}}/k_{\mathrm{B}} = d(\ln \rho)/\mathrm{d}(1/T)$, gives a strongly temperature-dependent $E_{\mathrm{eff}}/k_{\mathrm{B}}$ rather than a constant activation scale.

We also carried out a Zabrodskii analysis using $W(T) = -d(\ln \rho)/\mathrm{d}(\ln T)$. For $\rho = \rho_0 \exp[(T_0/T)^p]$, ln $W$ should be linear in ln(1/$T$), with slope $p$. No robust linear regime with $p$ = 1/4, 1/3, 1/2, or 1 is observed. The $p$ = 1/4 and $p$ = 1/3 limits correspond to three-dimensional and two-dimensional Mott variable-range hopping, respectively [24]. The $p$ = 1/2 form is associated with Efros-Shklovskii variable-range hopping in the presence of a Coulomb gap [25]. The derivative method used to test these hopping forms follows the Zabrodskii analysis [26]. The transport therefore does not support a unique assignment to conventional Mott variable-range hopping, Efros-Shklovskii variable-range hopping, or simple Arrhenius transport. The essential result is nonmetallic apparent pellet transport with no single activation or hopping scale over the measured temperature range. Because the measurement was performed on compacted pellets, we do not extract a unique bulk charge gap.

Figure 2(b) shows the low-temperature heat capacity plotted as $C/T$ versus $T^2$. The data are described by $C/T = \gamma + \beta T^2$, giving $\gamma$ = 6.88(7) mJ mol$^{-1}$ K$^{-2}$ and $\beta$ = 0.352(2) mJ mol$^{-1}$ K$^{-4}$. Using $\theta_{\mathrm{D}} = [12\pi^4 nR/(5\beta)]^{1/3}$, with the number of atoms per formula unit $n$ = 5 for $CdRuO_3$, we obtain $\theta_{\mathrm{D}}$ = 302(1) K. The finite residual $\gamma$ term indicates a residual low-temperature heat-capacity contribution and is incompatible with the simplest phonon-only hard-gap limit. However, in the absence of field-dependent low-temperature heat-capacity data, this term cannot be assigned uniquely to itinerant electronic quasiparticles. Possible contributions include residual electronic states, dilute magnetic defects, disorder-related low-energy excitations, or a combination of these effects.

Together with the nonmetallic and non-Arrhenius apparent pellet transport, the finite residual linear heat-capacity coefficient points to a non-ideal low-temperature state rather than to either a conventional good metal or a simple clean activated insulator. This conclusion is deliberately limited: the present data indicate correlation-sensitive low-energy behavior, but they do not determine a unique bulk charge gap.

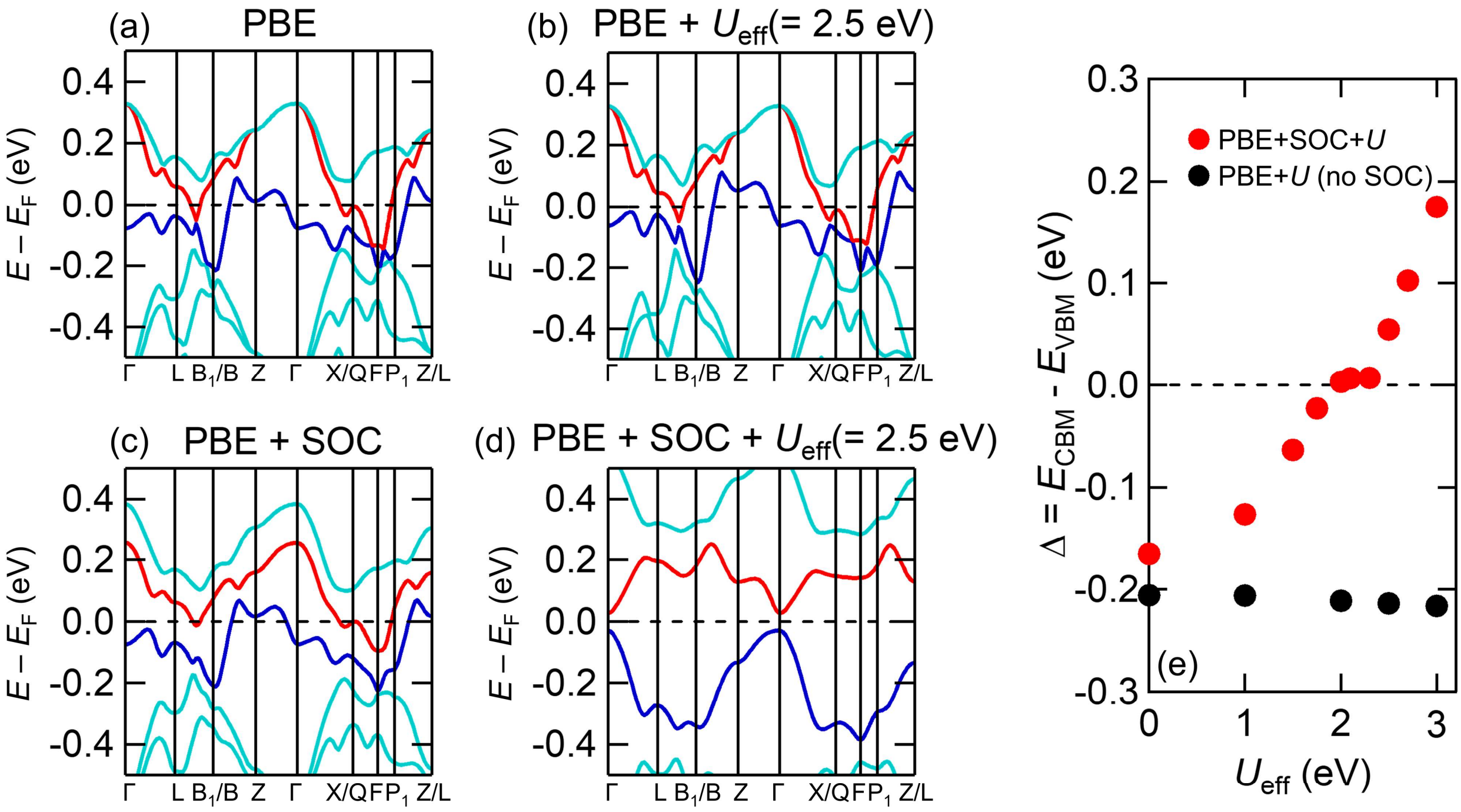

**Fig. 3** Spin-orbit and correlation dependence of the nonmagnetic electronic structure of $CdRuO_3$. (a) PBE band structure without spin-orbit coupling at $U_{eff} = 0$. (b) PBE+$U$ band structure without spin-orbit coupling at $U_{eff} = 2.5$ eV. (c) PBE+SOC band structure at $U_{eff} = 0$. (d) PBE+SOC+$U$ band structure at $U_{eff} = 2.5$ eV. The nominal highest occupied and lowest unoccupied bands used to define Δ are highlighted in blue and red, respectively; adjacent bands are shown in cyan. The high-symmetry labels follow the Setyawan–Curtarolo convention for the rhombohedral primitive cell. (e) $U_{eff}$ dependence of the signed indirect gap $\Delta = E_{CBM} - E_{VBM}$, evaluated on an 18 × 18 × 18 uniform $k$-point mesh. Positive and negative values indicate an insulating gap and an indirect band overlap, respectively. The no-SOC calculations remain metallic or semimetallic up to $U_{eff} = 3$ eV, whereas the SOC calculations approach the gap-closing condition near $U_{eff} \approx 2$ eV and develop a clearly positive gap at larger $U_{eff}$. Complete $U_{eff}$-dependent band structures and $k$-point convergence tests are presented in the Supplemental Material.

### 3. Electronic structure

Figure 3 compares representative nonmagnetic band structures with and without spin-orbit coupling. At $U_{eff} = 0$, both PBE and PBE+SOC give bands crossing the Fermi level, consistent with the metallic reference state previously predicted for $CdRuO_3$ [13].

The metallic or insulating character was evaluated using the signed indirect gap $\Delta = E_{CBM} - E_{VBM}$, where $E_{VBM}$ and $E_{CBM}$ were obtained from the extrema of the nominal valence and conduction bands on uniform $k$-point meshes. Without spin-orbit coupling, increasing $U_{eff}$ produces only minor changes in the low-energy band dispersions. The signed indirect gap remains negative throughout $0 \leq U_{eff} \leq 3$ eV. Its magnitude is approximately −0.20 to −0.21 eV and becomes slightly more negative with increasing $U_{eff}$. The PBE+U solutions therefore remain metallic or semimetallic over the investigated range and show no tendency toward gap opening.

A qualitatively different evolution is obtained when spin-orbit coupling is included. At $U_{eff} = 0$, spin-orbit coupling reduces the magnitude of the indirect band overlap but does not open a gap. With increasing $U_{eff}$, however, Δ increases rapidly and approaches zero around $U_{eff} \approx 2.0$ eV. Because the calculated Δ near $U_{eff} = 2.0$ eV is comparable to the numerical resolution, this point is described as being close to the gap-closing condition rather than as a well-established insulating state. At $U_{eff} = 2.5$ eV, the PBE+SOC+$U$ calculation gives a clearly positive gap of approximately 55 meV. The valence-band maximum and conduction-band minimum both occur at Γ, indicating a direct gap.

At the same $U_{eff} = 2.5$ eV, the corresponding PBE+$U$ calculation without spin-orbit coupling retains an indirect band overlap of approximately 0.21 eV. Thus, neither spin-orbit coupling alone nor $U_{eff}$ alone produces an insulating solution over the physically motivated parameter range examined here. The insulating solution emerges from their cooperative action.

The contrasting $U_{eff}$ dependences are summarized in Fig. 3(e). The 18 × 18 × 18 and 20 × 20 × 20 calculations produce nearly indistinguishable band dispersions and signed gaps for the selected SOC calculations at $U_{eff} = 1.5$, 2.0, 2.5, and 3.0 eV.

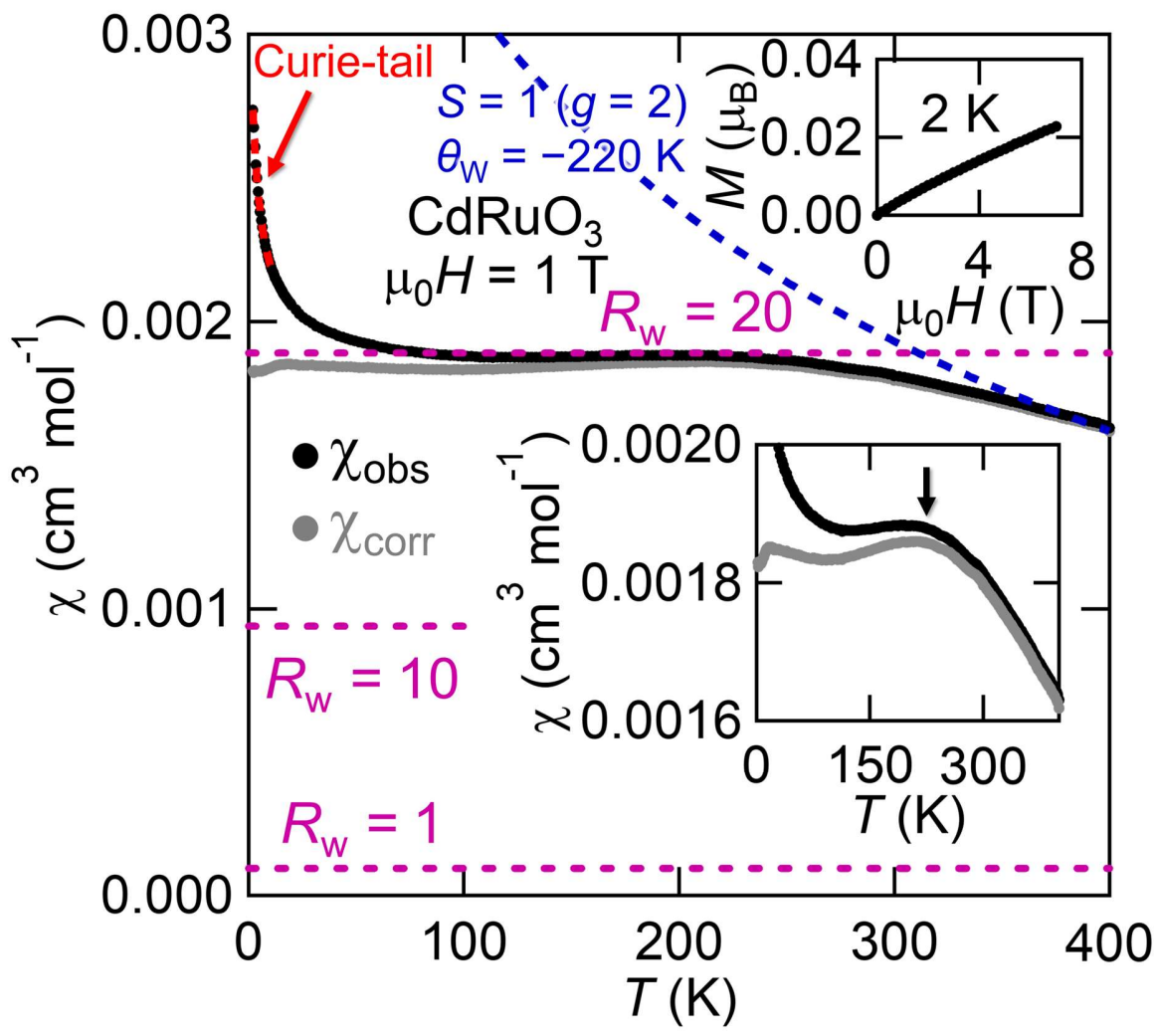

**Fig. 4** Temperature dependence of the magnetic susceptibility measured under $\mu_0H = 1$ T. Filled black symbols show the measured susceptibility $\chi_{obs}$, defined as the total measured susceptibility before any subtraction; it contains both the intrinsic response of $CdRuO_3$ and the low-temperature Curie-tail contribution from dilute paramagnetic defects. Gray symbols show the defect-corrected susceptibility $\chi_{corr} = \chi_{obs} - \chi_{imp}$, where $\chi_{imp}$ is the fitted impurity contribution The red solid line represents the Curie-Weiss-type impurity contribution fitted below 20 K with $C_{imp} = 4.80 \times 10^{-3}$ cm$^3$ K mol$^{-1}$. This contribution was subtracted from $\chi_{obs}$ to obtain $\chi_{corr}$. The fitted Curie constant corresponds to an effective impurity moment of 0.196 $\mu_B$ per formula unit. The upper inset shows the magnetization curve measured at 2 K; the small and nearly linear *M-H* response up to 7 T indicates the absence of a sizable ferromagnetic impurity contribution. The lower inset shows an expanded view of the susceptibility in the intermediate- and high-temperature region, where the arrow marks the broad maximum or shoulder around 180-220 K. The magenta dashed lines indicate Pauli susceptibilities estimated from $\gamma = 6.88(7)$ mJ mol$^{-1}$ K$^{-2}$ for Wilson ratios $R_W = 1$, 10, and 20, under the conditional assumption that $\gamma$ is entirely due to ordinary quasiparticles. The $R_W = 1$ value is $\chi_{Pauli} = 9.4 \times 10^{-5}$ cm$^3$ mol$^{-1}$. The blue dashed curve represents a representative spin-only $S = 1$ Curie-Weiss response with $C = 1.00$ cm$^3$ K mol$^{-1}$ for $g = 2$ and $\theta_W = -220$ K. The comparison shows that $\chi_{corr}$ is not naturally described by either an ordinary Pauli metal or independent spin-only $S = 1$ $Ru^{4+}$ moments.

The apparently large difference between the PBE+SOC and PBE+SOC+$U$ band structures does not represent a discontinuous change of crystallographic symmetry. In the rotationally invariant Hubbard scheme, the correction is determined by the self-consistent local Ru 4d occupation matrix and is therefore orbital dependent. The weak $U_{eff}$ dependence without SOC indicates that the Hubbard potential acts almost uniformly on the relevant low-energy $t_{2g}$ states in that solution. SOC lifts part of the degeneracy and mixes orbital and spin characters, allowing the Hubbard correction to shift the resulting spin-orbit-entangled states non-uniformly. Since their orbital compositions vary with $k$, the reconstruction is also $k$ dependent.

Figure S1 shows that the low-energy bands evolve continuously with $U_{eff}$. During this evolution, avoided crossings exchange the energy ordering of individual bands. The numerical band indices used in the plots are assigned according to energy at each k point and are therefore not conserved labels across different $U_{eff}$ values. Consequently, the $U_{eff} = 0$ and $U_{eff} = 2.5$ eV bands need not admit a simple one-to-one correspondence by band index, even though the low-energy $t_{2g}$-derived subspace evolves continuously. The insulating gap appears when this continuous SOC–U reconstruction removes the indirect band overlap.

These results demonstrate that spin-orbit coupling does not act merely as a small rigid shift of the band energies. Rather, it strongly enhances the response of the non-cubic Ru $t_{2g}$ manifold to local Coulomb interaction. We therefore describe the calculated reconstruction as spin-orbit-enhanced correlation sensitivity. The calculations do not establish a pure $J_{eff}$ state, a unique orbital-ordering pattern, or a specific many-body gap mechanism.

Complete $U_{eff}$-dependent band structures and convergence tests are provided in the Supplemental Material.

## 4. Magnetic susceptibility

Figure 4 shows the total magnetic susceptibility $\chi_{obs}(T)$ measured under $\mu_0H = 1$ T. The measured susceptibility contains a Curie-like upturn at low temperatures. To isolate this contribution, the upturn below 20 K was fitted by a Curie-Weiss-type impurity term, $\chi_{imp}(T) = C_{imp}/(T - \theta_{imp})$, which represents a dilute paramagnetic-defect contribution. The fit gives $C_{imp} = 4.80 \times 10^{-3}$ cm$^3$ K mol$^{-1}$ and $\theta_{imp} = -3.297$ K, corresponding to an effective impurity moment of 0.196 $\mu_B$ per formula unit. The defect-corrected susceptibility was then defined as $\chi_{corr}(T) = \chi_{obs}(T) - \chi_{imp}(T)$. If this Curie term is modeled as dilute $S = 1$, $g = 2$ moments, the fitted Curie constant corresponds to approximately 0.48 mol% of such moments; if modeled as dilute $S = 1/2$, $g = 2$ moments, it corresponds to approximately 1.28 mol%. Here, the notation $\chi_{imp}$ does not imply the identification of a specific impurity phase; it denotes the phenomenological Curie–Weiss contribution associated with dilute paramagnetic defects.

The low-temperature term is assigned to dilute paramagnetic defects for three reasons. First, it is confined to the low-temperature region and can be separated from the broad intermediate-temperature response. Second, the inferred defect concentration is far too small to represent the bulk Ru sublattice. Third, the magnetization at 2 K remains small and nearly linear up to 7 T, excluding a sizable ferromagnetic impurity contribution. Possible microscopic sources include local Li retention below the EDS detection limit, oxygen nonstoichiometry, cation disorder, grain-boundary defects, or locally disrupted exchange paths introduced during the topochemical reaction. Sample-dependent low-temperature

Curie tails have also been reported in the topochemically prepared honeycomb ruthenate $Ag_3LiRu_2O_6$ [8].

After subtracting $\chi_{imp}$, the defect-corrected susceptibility $\chi_{corr}$ remains of order $10^{-3}$ $cm^3$ $mol^{-1}$. It shows a weakly nonmonotonic temperature dependence: a broad maximum or shoulder appears around 180–220 K and becomes nearly temperature independent on cooling toward approximately 100 K.

An ordinary Pauli response is insufficient to explain its magnitude. If the measured γ were attributed entirely to itinerant quasiparticles with Wilson ratio $R_W = 1$, the estimated Pauli susceptibility would be 9.4 × $10^{-5}$ $cm^3$ $mol^{-1}$. Reproducing $\chi_{corr}$ solely within an itinerant quasiparticle picture would require an effective $R_W$ of approximately 17-20. Such a strongly enhanced itinerant response is not excluded, but it does not by itself explain the weakly nonmonotonic temperature dependence.

Independent spin-only $S$ = 1 $Ru^{4+}$ moments are also inconsistent with the data. The spin-only Curie constant for S = 1 and g = 2 is approximately 1.00 $cm^3$ K $mol^{-1}$. A representative Curie-Weiss response with this Curie constant is substantially more temperature dependent and remains monotonic in the paramagnetic regime.

The corrected susceptibility therefore lies between ordinary Pauli and independent-local-moment limits. A Van Vleck-type contribution may arise through field-induced mixing between nonmagnetic and magnetic states [27], but the present data do not establish a specific $J_{eff}$ level scheme or a spin-orbital-singlet ground state.

### D. Discussion

The experimental structure provides a direct test of the structural and electronic contrast predicted for $MgRuO_3$ and $CdRuO_3$ [13]. In the calculated $MgRuO_3$ branch, one Ru-Ru bond becomes strongly shortened, selecting one $t_{2g}$ orbital and splitting it into bonding and antibonding states. The Fermi surfaces are then formed primarily by the remaining two $t_{2g}$ orbitals. In the calculated $CdRuO_3$ branch, the larger A-site ion suppresses this bond-selective instability, the structure remains close to $R\overline{3}$, and the three $t_{2g}$ orbitals contribute almost equally to the metallic Fermi surfaces. The experimental $CdRuO_3$ structure confirms this non-dimerized branch: all nearest-neighbor Ru-Ru bonds are crystallographically equivalent, and their 3.076 Å length is far longer than the short molecular-orbital dimer bond in $Li_2RuO_3$.

The absence of Ru-Ru dimerization does not imply an undistorted Ru environment. $CdRuO_3$ retains a large local $RuO_6$ distortion while preserving equivalent Ru-Ru links. The dominant lattice perturbation in $CdRuO_3$ is therefore qualitatively different from the bond-selective distortion of $MgRuO_3$ or $Li_2RuO_3$. In those systems, a selected Ru-Ru bond removes one orbital channel through bonding-antibonding splitting. In $CdRuO_3$, the three Ru-Ru directions remain equivalent, while the local non-cubic crystal field reconstructs the underlying $t_{2g}$ manifold.

The matched calculations show that this non-cubic manifold responds qualitatively differently to $U_{eff}$ depending on spin-orbit coupling. Without spin-orbit coupling, the indirect band overlap remains close to 0.2 eV and becomes slightly larger as $U_{eff}$ increases to 3 eV. The strong structural distortion and $U_{eff}$ alone therefore do not destabilize the spin-unpolarized metallic or semimetallic solution over the investigated range. With spin-orbit coupling, the same bands become strongly $U_{eff}$ dependent, and a positive gap develops at moderate $U_{eff}$.

The strong band reconstruction induced by $U_{eff}$ in the SOC calculation does not require additional crystallographic symmetry lowering. Spatial symmetry constrains the allowed degeneracies and hybridizations but does not fix the relative energies or dispersions of the symmetry-allowed states. In the rotationally invariant PBE+$U$ treatment, the Hubbard potential is determined by the self-consistent spin-orbital occupation matrix rather than by spatial symmetry alone. Without SOC, the correction leaves the relative low-energy dispersions nearly unchanged. Once SOC has differentiated and mixed the t2g states, the same local interaction produces state- and $k$-dependent shifts, avoided crossings, and band reordering. The resulting gap should therefore be understood as a cooperative self-consistent reconstruction of the SOC-entangled, non-cubic $t_{2g}$ manifold, rather than as a rigid Hubbard splitting of three independently identifiable bands.

The calculated gap is consequently cooperative rather than attributable to one interaction alone. The non-cubic crystal field defines the underlying orbital environment, spin-orbit coupling reorganizes the orbital and spin character near the Fermi level, and local Coulomb interaction amplifies this reconstruction into an insulating solution. This mechanism is distinct from the bond-selective molecular-orbital formation predicted for $MgRuO_3$. It should also not be described as a pure $J_{eff} = 0$ state because the large non-cubic field mixes the ideal $J_{eff}$ sectors. In addition, the present static mean-field calculations do not determine a unique orbital-ordering pattern. Possible lower-symmetry orbital-polarized no-SOC solutions were not systematically explored.

The transport and heat-capacity measurements are qualitatively consistent with instability of the itinerant reference state but do not establish a homogeneous bulk gap. The compacted-pellet resistivity is nonmetallic, yet neither a single Arrhenius scale nor a standard variable-range-hopping exponent describes the full temperature range. Grain-boundary barriers, porosity, and intergrain tunneling may contribute. Conversely, the residual linear heat-capacity coefficient indicates low-energy spectral weight, but its origin may include defect- or disorder-related contributions. The discrepancy between nonmetallic pellet transport and metallic PBE+SOC bands should therefore not be interpreted as direct evidence of a sharp thermodynamic metal-insulator transition. Instead, it demonstrates that the actual low-energy response is more correlation and sample sensitive than the uncorrelated metallic reference state.

The susceptibility provides an independent diagnostic. A robust weakly correlated multiorbital metal would be expected to show a predominantly Pauli-like response. The measured $\chi_{corr}$ is much larger than the $R_W$ = 1 Pauli estimate and would require an effective $R_W$ close to 20 if assigned entirely to itinerant quasiparticles. The opposite spin-only limit is also inadequate because independent $S$ = 1 moments would produce a much stronger and monotonic Curie-Weiss response. The observed magnetic response therefore lies between ordinary itinerant and independent-local-moment limits.

Spin-orbit coupling is not required to explain the dilute low-temperature Curie tail, but it remains relevant to the intrinsic susceptibility because it modifies the orbital character and magnetic matrix elements of the low-energy bands. A Van Vleck-like interband contribution is therefore possible in the strongly non-cubic $d^4$ environment. The present data cannot, however, distinguish such a contribution from strongly enhanced itinerant spin susceptibility or other correlation-induced effects. NMR, μSR, optical conductivity, photoemission, and inelastic spectroscopy will be required to resolve the microscopic excitation spectrum.

The principal significance of $CdRuO_3$ can therefore be stated without assigning a specific insulating or $J_{eff}$ ground state. The material experimentally realizes the non-dimerized structural branch predicted for ruthenium ilmenites, but the associated multiorbital metallic state is not robust. Suppression of Ru-Ru molecular-orbital dimerization leaves a strongly distorted $t_{2g}$ manifold whose response to local Coulomb interaction is greatly amplified by spin-orbit coupling. $CdRuO_3$ thus provides a route to electronic reconstruction in a honeycomb ruthenate that does not rely on Ru-Ru bond-selective dimer formation.

## E. Conclusions

$CdRuO_3$ forms a crystallographically non-dimerized $R\overline{3}$ ilmenite structure with equivalent Ru-Ru bonds and a strongly distorted $RuO_6$ honeycomb layer. The experimental structure confirms the non-dimerized branch previously predicted for $CdRuO_3$, but the corresponding metallic state is not robust against combined spin-orbit coupling and local Coulomb interaction. PBE+$U$ calculations without spin-orbit coupling remain metallic or semimetallic throughout $0 \leq U_{eff} \leq 3$ eV, whereas PBE+SOC+$U$ evolves from an indirect band overlap to a direct gap of approximately 55 meV at $U_{eff} = 2.5$ eV. Spin-orbit coupling therefore strongly enhances the correlation sensitivity of the non-cubic Ru $t_{2g}$ manifold. The apparent pellet resistivity is nonmetallic but does not define a unique conduction law, and the residual heat-capacity term indicates unresolved low-energy spectral weight. After subtraction of a dilute Curie-like defect contribution, the susceptibility is incompatible with both an ordinary Pauli response and independent spin-only $S = 1$ $Ru^{4+}$ moments. $CdRuO_3$ thus provides a non-dimerized $4d^4$ ilmenite platform in which local lattice distortion, spin-orbit coupling, electronic correlation, and itinerancy compete without Ru-Ru molecular-orbital dimerization.

## Acknowledgments

We thank A. Yamamoto for assistance with high-pressure pellet preparation for the resistivity measurements at the Shared Equipment Centers of Shibaura Institute of Technology. This work was supported by JSPS KAKENHI Grant Nos. JP19K14646, JP21K03441, JP25K01496, JP23H04616, JP25H01649, JP25H01403, and JP22K14002, and by JST PRESTO Grant No. JPMJPR23Q8. Part of this work was carried out under the joint-use program of the Institute for Solid State Physics, the University of Tokyo (Project Nos. 202311-MCBXG-0021, 202311-MCBXG-0025, 202406-MCBXG-0100, 202406-GNBXX-0095, 202406-MCBXG-0101, 202411-MCBXG-0033, and 202411-MCBXG-0034).

## Data availability

The data that support the findings of this article are openly available [28].

**Supplemental Materials:**
**"Spin-orbit-enhanced correlation sensitivity and anomalous magnetic response in non-dimerized $4d^4$ ilmenite $CdRuO_3$"**

## S1. Geometrical parameters of the refined $RuO_6$ octahedron

Local bond lengths and bond angles around Ru were calculated from the refined average $R\bar{3}$ structure of $CdRuO_3$ using the crystallographic information file. The refined $RuO_6$ octahedron contains three short Ru-O bonds and three long Ru-O bonds. The Ru-O distances are 1.9055 Å × 3 and 2.1279 Å × 3, giving an average Ru-O distance of 2.0167 Å. The normalized bond-length variance was calculated as

$$\Delta_d = \frac{1}{6}\sum_i [(d_i - d_{avg})/ d_{avg}]^2,$$

where $d_i$ is an individual Ru-O bond length and $d_{avg}$ is the average Ru-O bond length. This gives $\Delta_d = 3.04\times10^{-3}$. The unnormalized bond-length variance is 0.01236 $Å^2$.

The O-Ru-O angles separate into twelve cis angles and three trans angles. The twelve cis angles are 76.180 deg × 3, 80.758 deg × 3, 99.222 deg × 3, and 103.744 deg × 3. The three trans angles are 175.305 deg × 3. The bond-angle variance was calculated from the twelve cis O-Ru-O angles as

$$\sigma^2 = \sum_i (\theta_i - 90)^2/11,$$

where $\theta_i$ denotes one of the twelve cis O-Ru-O angles. The resulting value is $\sigma^2 = 150.10$ $deg^2$, corresponding to an RMS angular deviation of 12.25 deg from an ideal octahedron.

Each oxygen atom bridges two nearest-neighbor Ru atoms in the honeycomb layer. The corresponding Ru-O-Ru angle is 99.242 deg, with Ru-O distances of 1.9055 Å and 2.1279 Å to the two Ru sites. The nearest-neighbor in-plane Ru-Ru distance in the honeycomb layer is 3.0760 Å, consistent with the absence of a short Ru-Ru dimer bond in the average structure. The geometrical layer repeat distance $c/3$ is 4.9528 Angstrom; the shortest interlayer Ru-Ru contact generated from the same average structure is 4.9360Å.

**Supplementary Table S1.** Local structural parameters of $CdRuO_3$ derived from the refined average $R\bar{3}$ structure. All distances are in Angstrom and all angles are in degrees. Multiplicities for Ru-O and O-Ru-O entries are given per $RuO_6$ octahedron. Multiplicities for Ru-O-Ru entries are given per bridging oxygen. Ru-Ru multiplicities are given per Ru site unless otherwise stated.

| **Quantity** | **Value** | **Multiplicity** |
|---|---|---|
| short Ru-O distance | 1.9055 Å | 3 |
| long Ru-O distance | 2.1279 Å | 3 |
| Average Ru-O distance | 2.0167 Å | - |
| Ru-O bond-length variance, unnormalized | 0.01236 Å$^2$ | - |
| Ru-O normalized bond-length variance $\Delta_d$ | $3.04 \times 10^{-3}$ | - |
| O-Ru-O cis angle | 76.180º | 3 |
| O-Ru-O cis angle | 80.758º | 3 |
| O-Ru-O cis angle | 99.222º | 3 |
| O-Ru-O cis angle | 103.744º | 3 |
| O-Ru-O trans angle | 175.305º | 3 |
| O-Ru-O bond-angle variance sigma2 | 150.10 deg$^2$ | - |
| RMS angular deviation | 12.25º | - |
| Ru-O-Ru angle | 99.242º | 1 per O |
| Ru-O distances for Ru-O-Ru bridge | 1.9055 Å, 2.1279 Å | 1 each |
| Ru-Ru nearest-neighbor distance | 3.0760 Å | 3 |
| Ru-Ru next-nearest in-plane distance | 5.3276 Å | 6 |
| Shortest interlayer Ru-Ru contact | 4.9360 Å | 1 |
| Geometrical Ru-layer repeat distance | 4.9528 Å | - |

## S2. Complete $U_{eff}$-dependent band structures

Figure S1 presents the complete low-energy band-structure series used to establish the contrasting $U_{eff}$ dependences with and without spin-orbit coupling. The no-SOC series includes PBE and PBE+$U$ calculations at $U_{eff}$ = 1, 2, 2.5, and 3 eV. The SOC series includes PBE+SOC and PBE+SOC+$U$ calculations at $U_{eff}$ = 1, 1.5, 1.75, 2.0, 2.1, 2.3, 2.5, 2.7, and 3.0 eV.

Without spin-orbit coupling, the low-energy band dispersions change only weakly and retain an indirect band overlap throughout the investigated range. With spin-orbit coupling, the bands near the Fermi level are progressively reconstructed as Ueff increases. The indirect band overlap approaches zero near $U_{eff} \approx 2$ eV, and a clearly positive gap is obtained at $U_{eff}$ = 2.5 eV.

## S3. *k*-point convergence of the band dispersions

To test the k-point convergence of the revised PBE+SOC+$U$ calculations, calculations at $U_{eff}$ = 1.5, 2.0, 2.5, and 3.0 eV were repeated using a 20 × 20 × 20 k-point mesh. All other calculation parameters, including the experimental structure, pseudopotentials, kinetic-energy cutoffs, smearing width of 0.001 Ry, and self-consistency threshold, were kept unchanged.

Figure S2 overlays the band structures obtained using 18 × 18 × 18 and 20 × 20 × 20 meshes. The dispersions are essentially indistinguishable throughout the displayed energy range, including the Ru $t_{2g}$ bands close to the Fermi level. The high-symmetry-path comparison is used only to demonstrate stability of the plotted dispersions; the signed gap values were evaluated from uniform-mesh eigenvalues as described in the main manuscript.

## S4. *k*-point convergence of the signed indirect gap

Figure S3 compares the signed indirect gap $\Delta = E_{CBM} - E_{VBM}$ obtained using 18 × 18 × 18 and 20 × 20 × 20 meshes. The comparison focuses on the SOC calculations close to and above the gap-closing condition, where the *k*-point sensitivity is most relevant.

The two meshes give consistent values and preserve the qualitative classification of the representative states. The $U_{eff}$ = 2.0 eV SOC solution remains close to the gap-closing condition, while the positive gaps at $U_{eff}$ = 2.5 and 3.0 eV remain stable. The no-SOC 18 × 18 × 18 results remain approximately −0.2 eV from $U_{eff}$ = 0 to 3 eV and are therefore far from the gap-closing condition.

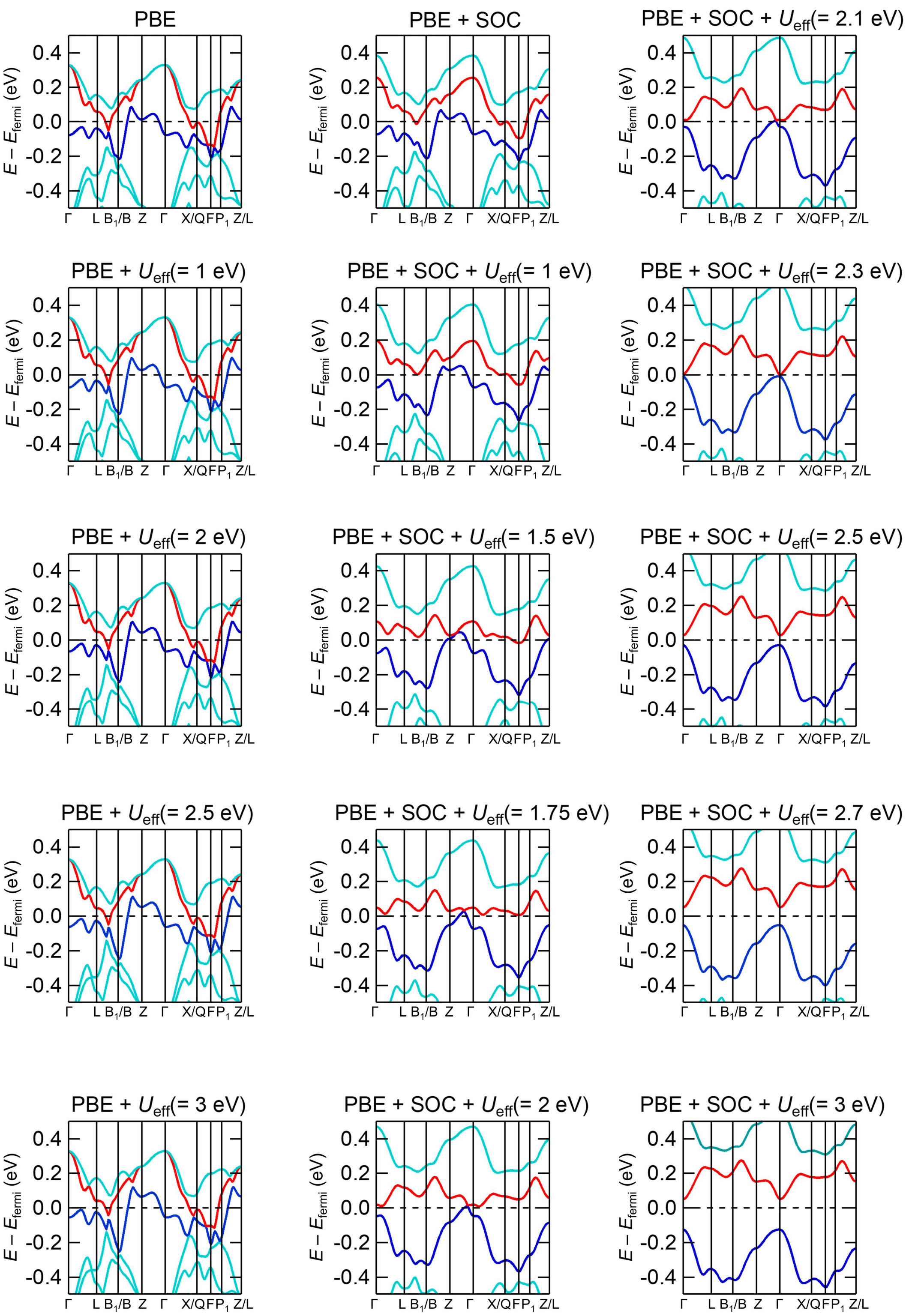

**Figure S1** Complete $U_{eff}$ evolution of the nonmagnetic electronic band structures of $CdRuO_3$ with and without SOC. The sequence demonstrates that the low-energy $t_{2g}$-derived band subspace evolves continuously as $U_{eff}$ increases. In the SOC calculations, avoided crossings and associated band reordering make a simple one-to-one correspondence between energy-ordered bands at $U_{eff} = 0$ and $U_{eff} = 2.5$ eV inappropriate. The band indices in each panel are assigned according to energy at each $k$ point and should not be interpreted as conserved labels across $U_{eff}$. No discontinuous crystallographic symmetry change is involved. The indirect band overlap is progressively removed through the continuous SOC–U reconstruction, and an insulating gap develops at moderate $U_{eff}$.

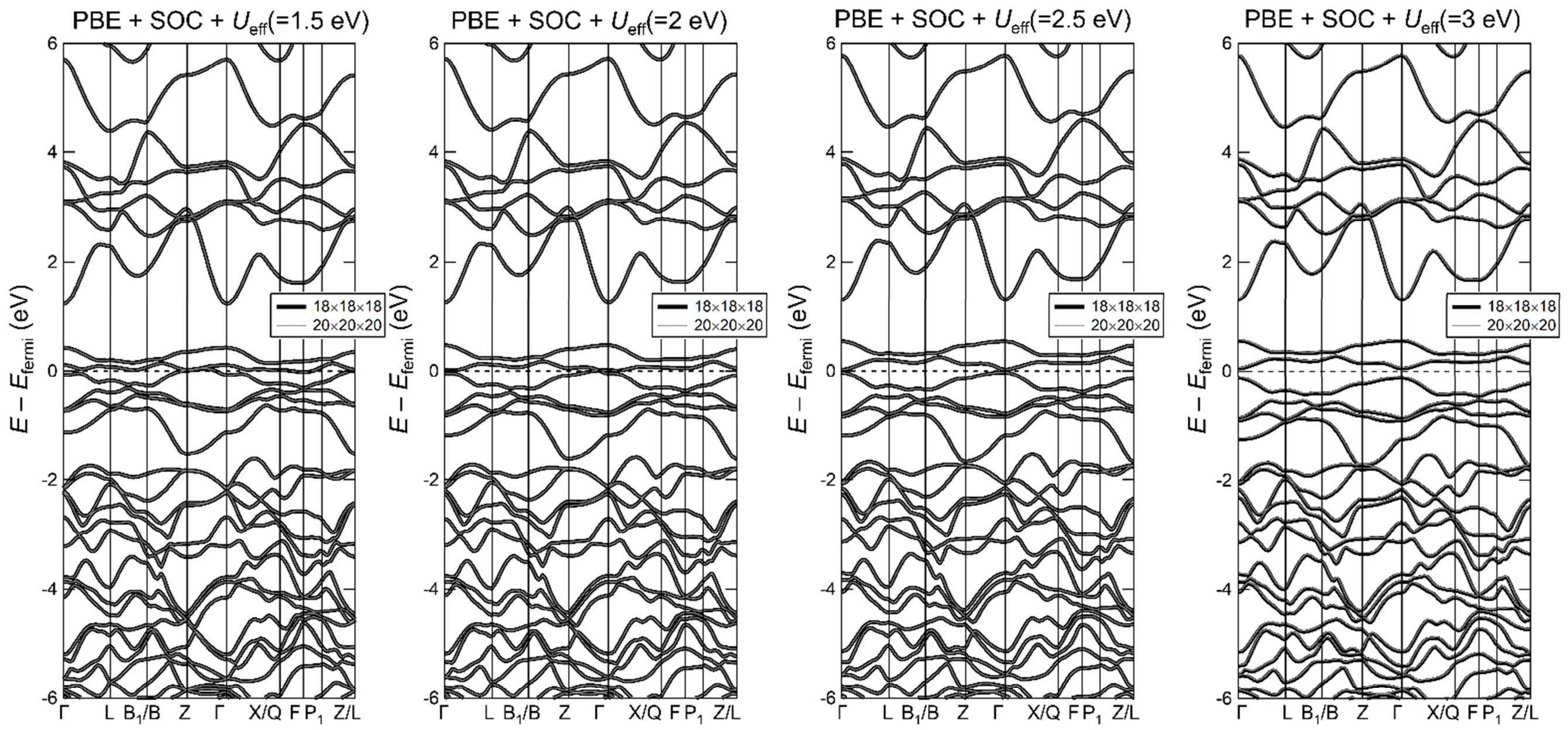

**Figure S2** Comparison of the nonmagnetic PBE+SOC+$U$ band structures obtained using 18 × 18 × 18 and 20 × 20 × 20 $k$-point meshes for $U_{\mathrm{eff}} = 1.5$, 2.0, 2.5, and 3.0 eV. Black and gray curves denote the 18 × 18 × 18 and 20 × 20 × 20 results, respectively. All other calculation parameters, including the experimental structure, pseudopotentials, kinetic-energy cutoffs, smearing width of 0.001 Ry, and self-consistency threshold of $1 \times 10^{-9}$ Ry, were kept identical. The band-path comparison demonstrates convergence of the plotted dispersions; the signed indirect gaps were evaluated independently from uniform-mesh eigenvalues.

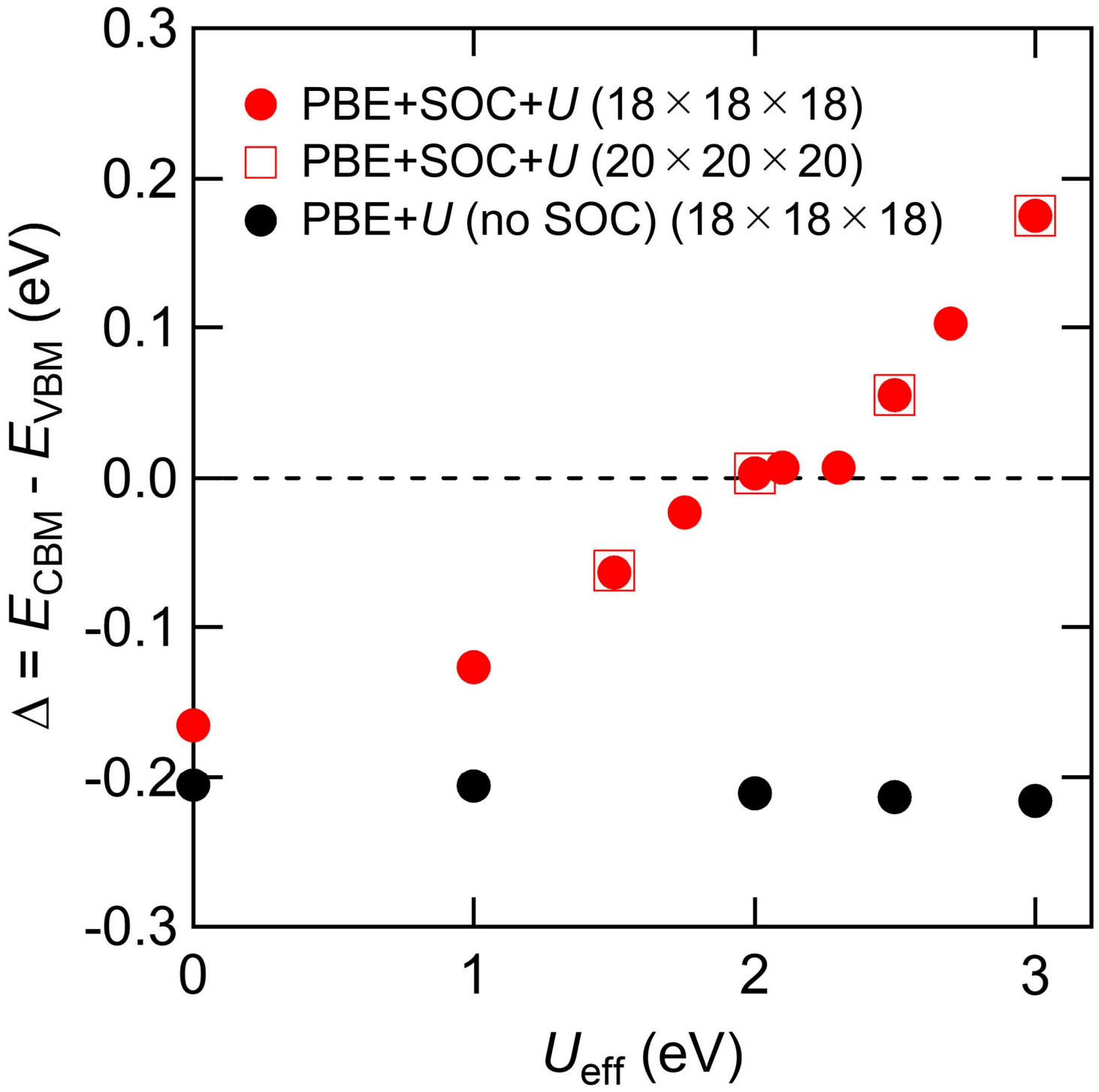

**Figure S3** $U_{\mathrm{eff}}$ dependence of the signed indirect gap $\Delta = E_{\mathrm{CBM}} - E_{\mathrm{VBM}}$. Filled red circles and open red squares denote PBE+SOC+$U$ results obtained using 18 × 18 × 18 and 20 × 20 × 20 k-point meshes, respectively. Filled black circles denote the PBE+$U$ results without spin-orbit coupling obtained using the 18 × 18 × 18 mesh. Positive and negative $\Delta$ values indicate an insulating gap and an indirect band overlap, respectively. The slight graphical overlap of the 18 × 18 × 18 and 20 × 20 × 20 symbols reflects the small numerical difference between the two meshes.